# Improving Deep Reinforcement Learning Agent Trading Performance in Forex using Auxiliary Task

Sahar Arabha, Davoud Sarani, and Parviz Rashidi-Khazaee

*Abstract*—**Advanced algorithms based on Deep Reinforcement Learning (DRL) have been able to become a reliable tool for the Forex market traders and provide a suitable strategy for maximizing profit and reducing trading risk. These tools try to find the most profitable strategy in this market by examining past market data. Artificial intelligent agents based on the Proximal Policy Optimization (PPO) algorithm, one of the DRL algorithms, have shown a special ability to determine a profitable strategy. In this research, to increase profitability and determine the optimal strategy for the PPO, an auxiliary task has been used. The auxiliary function helps the PPO to model the reward function in a better way by recognizing and classifying patterns to obtain additional information from the problem's inputs and data. The results of simulation and backtesting on the EUR/USD currency pair have shown that the new proposed model has been able to increase the overall return from -25.25% to 14.86% and the value of sharp ratio from -2.61 to 0.24 in dataset 1 and increased overall return from 2.12% to 42.22% and the value of sharp ratio from -2.93 to 0.47 in dataset 2. These results indicate better identification of trends, timely trading, and great improvement in risk reduction while increasing the profit. It has also reduced the transaction risk and can be used as an effective tool for automatic trading in the complex forex market**

## I. INTRODUCTION

Today, due to economic fluctuations, most people have become more willing to invest in the forex market to earn more money in this market. To be successful in this market, it is necessary to spend most of your time sitting behind the system so that you can find a suitable strategy with high profits. But the challenge is whether it is possible to find a strategy with suitable profits or not. Second, it is necessary to spend the whole day sitting behind the system and just trade? Today, researchers have shown that Deep Reinforcement Learning (DRL) extracts meaningful features from data and makes decisions based on them [1]. Also, it has been shown that artificial intelligent agents based on DRL have helped to solve this problem [2].

Deep Learning (DL) algorithms such as short-term memory networks (LSTM) have been used to predict and trade in the forex market. Tsantekidis et al. have used LSTM networks to design a model for executing optimal trades in the forex market [3]. To improve the performance of LSTM and Bidirectional LSTM (BI-LSTM), it is necessary to optimally and properly adjust their various parameters, and the accuracy of their prediction depends on the optimization mechanism of determining their parameters [4]. Liu et al. devised a strategy for automated Bitcoin transactions and designed an automated transaction execution strategy based on Proximal Policy Optimization (PPO) using LSTM as a basis for policy generation. Experimental results have shown that the LSTM algorithm has improved the performance of PPO [5].

The DRL methods, with special capabilities, have provided more possibilities and allowed more exploitation of time series data [6] - [7]. In recent years, published DRL algorithms such as Asynchronous Advantage Actor Critic (A3C) [8], Deep Q-Network (DQN) [9], and PPO [10], have been able to perform in various fields and have made it possible to perform optimal, profitable and low-risk transactions in trading markets [6]. If the DRL agent is properly trained through time series data, trends can be predicted correctly, and more profitable transactions can be executed according to market trends. These algorithms have been able to achieve significant results in trading markets [3].

Lele et al. used a Reinforcement Learning (RL) agent to learn stock market trends and suggest trading decisions that maximize profits. They used Trust Region Policy Optimization (TRPO) [11], PPO, and Vanilla (VPG) algorithms to increase the profit from trading shares of three different companies including Microsoft, Apple, and Nike. They created a trading agent, which performs trading actions on the environment and re-train according to the received rewards. The result showed that the TRPO algorithm performed better than the other two algorithms in terms of profit [12].

Lin et al. used an end-to-end framework to improve transaction execution quality and reduce per-transaction costs

This paragraph of the first footnote will contain the date on which you submitted your paper for review, which is populated by IEEE. It is IEEE style to display support information, including sponsor and financial support acknowledgment, here and not in an acknowledgment section at the end of the article. For example, "This work was supported in part by the U.S. Department of Commerce under Grant 123456." The name of the corresponding author appears after the financial information, e.g.

*(Corresponding author: Dr. Parviz Rashidi-Khazaee).*
Sahar Arabha is Mastery program student at Information Technology and Computer Engineering Department, Urmia University of Technology, Urmia, Iran (e-mail: sa.arabha@it.uut.ac.ir).

Davoud Sarani, is Mastery program student at Information Technology and Computer Engineering Department, Urmia University of Technology, Urmia, Iran (e-mail: sarani.davoud@it.uut.ac.ir).

Dr. Parviz Rashidi-Khazaee is Assistant Professor at Information Technology and Computer Engineering Department, Urmia University of Technology, 4km Band Road, Urmia, Iran (e-mail: p.rashidi@uut.ac.ir, pr.rashidi@gmail.com).

by optimizing the PPO algorithm using two separate neural network architectures: 1) LSTM and 2) Fully Connected Networks (FCNs) to be able to perform more efficient transactions. They used a distributed reward signal for their reward function and results have shown that the Implementation Shortfall (difference between the decision price and the final price of a transaction), standard deviation profit-loss ratio, and performance of the proposed model outperform Almgren and Chriss model, Volume-Weighted Average Price, Time-Weighted Average Price and DQN [13].

Tsai et al. addressed the long-term problem of unstable trends in DL predictions and used an RL method to make transactions. They encrypted their data with the Gramian Angular Field and designed their trading model with DQN and PPO and used it for trading of three different currency pairs (AUD/USD, GBP/USD, EUR/USD). The comparison result showed that The PPO algorithm was considered superior for optimizing forex trading strategies [7].

DRL methods try to find and apply the optimal trading strategy by using the reward received from the environment. Therefore, most of the research done in the field of the forex trading market is focused on how to shape the reward function. The reward function plays a pivotal role in increasing the agent's ability to identify and execute profitable transactions and thus improve overall performance. Konstantinos used an RL agent along with an auxiliary task to learn investment strategies. He used an agent whose purpose is to learn from historical data and suggest investment strategies for buying and selling to earn profit through changes in the price of digital currency. The main model uses LSTM networks for feature extraction and fully connected layers for the output function and auxiliary regression output. The Sharpe Ratio and Drawdown results indicate that the use of the auxiliary function gives much better results than the Base-line and Buy and Hold method [14]. Tsantekidis et al. addressed the problem of the noise of sparse bonuses and proposed the design of an additional bonus to increase the return and Sharpe Ratio in forex trading. By adding a trailing reward to the main reward Profit and Loss (PnL) and using a relatively more complex network structure consisting of LSTM and FC layers, they have helped to increase efficiency and the Sharpe Ratio of transactions. In addition, they used the Actor-Critic (AC) method for the value of the situation and the action performed in each situation, and finally, they used the PPO and Double DQN, to train the RL agent. They designed separate neural networks for each deep reinforcement learning algorithm. Their obtained results indicate that adding a additional rewards to the main rewards has increased the sharp ratio and return [3].

Lele et al. used TRPO, PPO, and Vanila to increase profits in the shares of 3 different companies [12]. Their reward function is not optimized for the risk measure. Also, they suggested to use a supervised learning model that takes sentiment analysis as input to predict stock trends to obtain more informed and efficient results [12]. Lin et al. used a distributed reward function and optimized the PPO algorithm for better execution quality in the stock market, which has brought favorable results [13].

In this research, the PPO algorithm has been used to find the optimal policy in the forex market so that the agent can adopt the best policy to perform profitable trading decisions. The problem that has been bothering the mind so far is that the reward received from this method has noise due to the dynamic environment of the forex market. In other words, the learning process is slow and unstable and cannot show the intended result [3]. Therefore, the main goal of this study is to build a new PPO-Based model by changing shaping the reward in such a way that it can have fast and sustainable results. The agent learns from various signals in the environment using pre-processed data based on unsupervised learning principles. This approach enables the agent to continuously increase his understanding without relying on direct rewards and thereby expand the scope of his learning capabilities. To do this, an auxiliary task was combined with PPO. The purpose of the auxiliary tasks is to recognize and classify patterns without human intervention, accelerate training, strengthen the agent, improve its overall performance, reduce the risk of transactions, and finally try to create a profitable model.

## II. BACKGROUNDS

At first, DL, RL, and DRL algorithms will be presented in this section briefly, and then the structure of the proposed model will be presented.

### A. Deep Learning

DL [1] - [2] is a subset of machine learning (ML), and research has shown that DL algorithms can outperform ML algorithms. There are several famous DL algorithms in this field such as RNN, LSTM, CNN, and Generative Adversarial Network. These algorithms have benefited from simulating the function of neurons in the human brain. While one neuron is not able to do complex tasks only. Therefore, the human brain has billions of neurons that are stacked in layers and form a network [1]. Based on that fact the structure of DL algorithms contains lots of layers and neurons.

### B. Reinforcement learning

RL stands as a distinct domain within the field of ML. Diverging from conventional ML approaches like supervised and unsupervised learning, RL entails agents learning by actively engaging with their environment, navigating through trial and error. RL has evolved rapidly over the past few years with a wide range of applications from building a recommendation system to self-driving cars. With the advent of new algorithms and libraries, RL stands out as one of the most promising domains within ML. Here we get to know some key elements of RL [1]:

**Agent:** An agent is a learning robot that makes intelligent decisions. For example, a chess player can be considered an agent because the chess player learns to make the best moves (decisions) to win the game.

**Environment:** The environment is the world in which the agent performs actions. For example, a chessboard is called an environment. Complex and high-dimensional environments can be challenging because the learning process may be slow or unstable.

**State and Action:** A state is a situation or moment in the environment that the agent can be in. E.g. in the game of chess, each position on the chessboard is called a state. The agent interacts with the environment and moves from one state to another by performing an action. E.g. in the chess game environment, the action is the move that the player (agent) makes.

**Rewards:** the reward is a value for performing an action, for example +1 for a good action and -1 for a bad action.

### C. Deep reinforcement learning algorithms

DRL integrates Deep Neural Networks with an RL framework to better handle high-dimensional input spaces such as images or raw data. These networks are used to approximate policy (agent strategy) or value functions (cumulative expected reward) for each state-action pair [1]. The goal of RL is to find the optimal policy, which is the policy that provides the maximum return. There are two methods to find the optimal policy: 1) value-based method and 2) policy-based method. In this paper, policy-based methods were used. Most policy-based methods use a stochastic policy, with a stochastic policy selects actions based on the probability distribution in the action space, which allows the agent to examine different actions instead of performing one action at a time. Several policy-based algorithms were available [15], in this paper, the PPO algorithm was used.

Before dealing with the PPO algorithm, it is necessary to have some detailed information about one of the popular policy-based methods for finding the optimal policy Actor-Critic (AC) [16].

### D. Actor-Critic

The AC is one of the most popular DRL methods and the PPO is designed based on AC. The AC method is a combination of value-based and policy-based optimal policy methods [16]. The Actor Critic framework comprises two distinct networks: 1) the Actor Network and 2) the Critic Network. The Actor Network is tasked with determining the most effective policy, whereas the Critic Network evaluates the policy generated by the Actor Network. The actor produces the optimal policy through the advantage function, through the advantage function actor can understand whether an action is really good or just gives it a similar value to all other actions, and the critic criticizes the actor's action in the acting point through the MSE function.

### E. Proximal Policy Optimization

PPO is an improved algorithm of TRPO and its implementation is simple. There are two different types of PPO algorithms: 1) PPO-clipped and 2) PPO-penalty. In this article, the PPO-clipped was used. To ensure that the policy updates are in the trust zone (the new policy is not far from the old policy), the PPO creates a function called the cut function, which ensures that the new policy is not far from the old policy.

The ratio of the new policy to the old policy is calculated through the following formula:

$$\underset{\theta}{maximize} \quad \mathbb{E}_t \left[ \frac{\pi_\theta(a_t|s_t)}{\pi_{\theta_{old}}(a_t|s_t)} A_t \right] \qquad (1)$$

$$L(\theta) = \mathbb{E}_t \left[ \frac{\pi_\theta(a_t|s_t)}{\pi_{\theta_{old}}(a_t|s_t)} A_t \right] \qquad (2)$$

Here, the expectation $\mathbb{E}_t[. . .]$ indicates the expected return average over a finite batch of samples. The $\pi$ indicates policy and the symbol $\theta$ represents parameters that need to be learned to improve the strategy. We parameterize the old policy with $\theta$ as $\pi_{\theta_{old}}$ and the new policy with $\theta$ as $\pi_\theta$.

Finally, considering the ratio of the new policy to the old policy as $r_t$, the following formula is obtained:

$$L(\theta) = \mathbb{E}_t[r_t(\theta) A_t] \qquad (3)$$

It means that it tries to maximize its policy along with the limitation. The objective function is as follows:

$$L(\theta) = \mathbb{E}_t[min(r_t(\theta) A_t, clip(r_t(\theta), 1 - \epsilon, 1 + \epsilon) A_t)] \qquad (4)$$

The $r_t$ is the ratio of the new policy to the old one, and $A_t$ is the advantage(benefit) function, which clipped $r_t$ in the range $[1 - \epsilon, 1 + \epsilon]$, but why? This can be explained by considering two cases of the benefit function, when the benefit is positive and when the benefit is negative.

#### 1) When the advantage is positive:

When $A_t > 0$, It signifies that the associated action ought to be favored above the mean of all alternative actions, so we can increase the value of $r_t(\theta)$ for that action so that it has a higher chance of being selected. However, while increasing $r_t$ ($\theta$) should not be increased too much to deviate from the old policy and clipped at $1 + \epsilon$.

#### 2) When the advantage is negative:

When $A_t < 0$, It signifies that the associated action ought to be favored above the mean of all alternative actions, so we can reduce the value of $r_t(\theta)$ for that action so that it has a lower chance of being selected, however, while reducing $r_t$ ($\theta$) should not be reduced too much to move away from the old policy and clipped at $1 - \epsilon$.

The problem that has been bothering the mind so far is that the reward received from PPO method has noise due to the dynamic environment of the forex market. In other words, the learning process is slow and unstable and cannot show the intended result. So a decision was made to change the way of shaping the reward in such a way that it can have fast and sustainable results.

Reward Shaping is a technique in DRL that involves modifying the reward function to guide agent learning toward a desired behavior [17]. A reward shaping model is shaping by providing additional information so that the agent can receive additional information about the environment as a reward, and since the higher the reward received, the greater the profit from the proposed model, so the learning agent will achieve advanced results by directly maximizing cumulative reward. In order to enhance learning efficiency, researchers have investigated the utilization of shaping incentives. Essentially, shaping involves enhancing the reward system to communicate existing knowledge to a standard RL system.

These artificial incentives assist in guiding the reinforcement learner towards favorable policies or steering away from unfavorable ones [18]. For example, the agent can be rewarded for detecting a change in market trends, or for identifying a particular pattern in the data. The reward shaping used in this research is based on a new shaping algorithm of DeepMind's article on Auxiliary tasks [19].

*F. Auxiliary Tasks*

Auxiliary tasks are additional cost functions that can help the RL agent as an additional leverage so that the agent itself can learn, predict, and observe data from the environment without human intervention. In other words, while DL algorithms have made considerable advancements across numerous fields, they demand expensive annotations on extensive datasets. Self-supervised learning (SSL) utilizing unlabeled data has emerged as a viable alternative by eliminating the need for manual annotation. SSL achieves this by crafting feature representations through auxiliary tasks that function without manual annotation. Consequently, models trained on these tasks can extract valuable latent representations, subsequently enhancing downstream tasks like object classification and detection[20]. This means that even in the absence of a strong reward signal, it measures and defines the loss (error or discrepancy) in a system using alternative or indirect indicators derived from unlabeled data. Unsupervised auxiliary tasks act as additional learning objectives alongside the main RL task. These auxiliary tasks are designed to help the agent learn useful features or representations of the data that can help improve the agent's performance on the main task. The main idea of combining RL with unsupervised auxiliary tasks is to exploit the data structure to accelerate and stabilize the learning process [21]. By forcing the agent to learn related features or patterns through auxiliary tasks, it can gain a better understanding of the environment and make more informed decisions in the main task. This approach can lead to faster convergence, improved efficiency, and better generalization when facing new situations. In general, RL with unsupervised auxiliary tasks is a promising way to advance the capabilities of AI agents and enable them to tackle complex real-world problems using reinforcement learning and unsupervised learning techniques.

## III. PROPOSED MODEL STRUCTURE

In this research, to correctly identify the trends and make profitable trades in the forex market, the performance of the reinforcement learning agent has been improved by using a separate auxiliary task. The main idea is to develop an agent that can learn from different signals in the forex market by recognizing patterns. This idea is derived from the logic of unsupervised learning, where the agent continues to develop its understanding of the environment even without direct reward. This approach expands the agent's learning scope and focuses on predicting and controlling various characteristics of the market environment and makes the agent able to flexibly control and understand his experiences by understanding the environment more.

Our contribution in this research is the use of auxiliary task to better and more optimally train the RL agent in such a way that it has better learning and makes more profitable trades. The auxiliary task makes the agent better understand the structure of the market environment and learn the optimal trading policies. For example, the agent learns that a certain set of situations leads to a suitable trading position. Therefore, it tries to learn them. The proposed auxiliary task examines the input data and after extracting the golden features, clusters them and creates a suitable label for each input data. This mechanism helps transform the problem from unsupervised learning to supervised learning using Forex time series data (features) and then the DRL agent uses the generated labels to recognize trends to make an optimal deal. To build the proposed model, the following steps are taken:

1) **Pre-processing stage**

   First, the time series of forex data is pre-processed and 5 new features are extracted.

2) **Labeling stage**

   5 features extracted in 16 previous time steps, a total of 80 features, are given to Auto Encoder (AE) to extract 12 golden features from its latent. Then the golden features are given to the K-Means algorithm to cluster each input instance into 12 separate clusters.

3) **Learning the optimal trading policy**

   The input data along with the created label are given to the Actor-Critic network so that the appropriate trading action is performed by the Actor network. The action performed by the actor is checked by the Critic and Auxiliary Task, and based on their feedback, the network is modified to learn a more optimal policy.

4) **Testing stage**

   The performance of the trained network on new data, the last seven months of the dataset, and their results are presented.

*A. Pre-Processing*

The dataset has OHLC features, which include hourly opening price, highest price, lowest price, and closing price of currency pairs. To improve the performance of the model and solve the time lag problem, as in [3], these features are processed and five new features are extracted as follows:

$$x1_t = \frac{p_c(t) - p_c(t-1)}{p_c(t-1)} \tag{5}$$

$$x2_t = \frac{p_h(t) - p_h(t-1)}{p_h(t-1)} \tag{6}$$

$$x3_t = \frac{p_l(t) - p_l(t-1)}{p_l(t-1)} \tag{7}$$

$$x4_t = \frac{p_h(t) - p_c(t-1)}{p_c(t-1)} \tag{8}$$

$$x5_t = \frac{p_c(t) - p_l(t-1)}{p_c(t-1)} \tag{9}$$

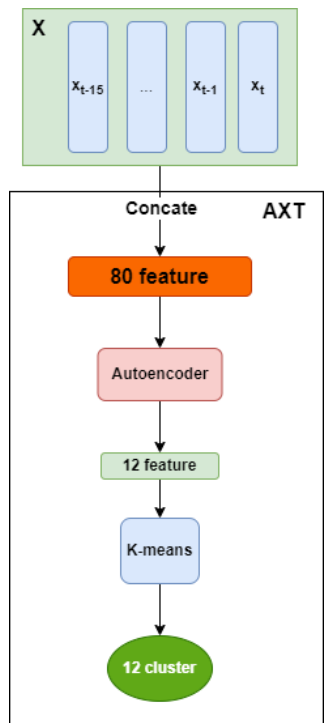

**Fig. 1**.The Proposed AXT Model Structure.

In which $p_c(t)$ is the closing price at time $t$, $p_h(t)$ is the highest price and $p_l(t)$ is the lowest price of the currency pair in the step $t$. $(t-1)$ indicates the previous time step.

The closing price has been used to calculate the return in a step-by-step manner, which is calculated as follows:

$$z_t = \frac{p_c(t) - p_c(t-1)}{p_c(t-1)} \qquad (10)$$

### B. Proposed Auxiliary Task Labeling Mechanism

One of the best ways to use the input data and help improve the performance of the model is to select or create optimal features from a set of features. Among the various methods, PCA is one of the effective techniques for extracting new and optimal features, which has been successfully used to extract superior features and optimize the effectiveness of stock market data clustering [22]. Another method is using Auto-Encoder (AE), which has special characteristics [23]. Initial evaluations have shown the performance of AE, therefore, it was used in this research, In addition to extracting new and optimal features, it also reduces computing costs [23]. In this step, the 5 features generated at a pre-processing step in the previous 16 time steps, a total of 80 features, are fed into the AE network. The Encoder part of the AE network has 4 layers and each layer has 80, 128, 64, and 32 neurons respectively. The AE network preprocesses the inputs and then 12 new features, called golden features, will be generated from the output of Encoder. Then the golden features are given to the K-Means algorithm for clustering and categorizing the trading data to produce a suitable label. The proposed structure for creating an appropriate label for input data shown in Fig. 1 named as AXT.

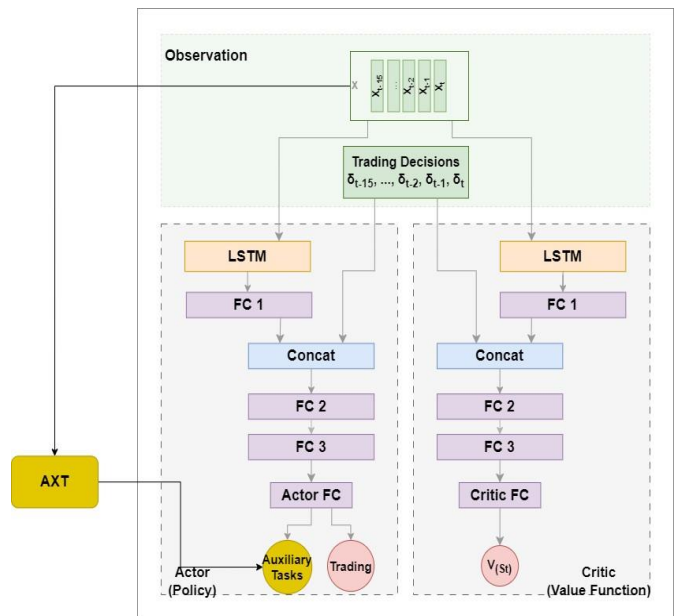

Fig. 2. Proposed trading agent structure

### C. The Proposed Structure of Trading Agent

Figure 2 shows the structure of the proposed trading agent architecture. In this structure, the input data is given to the ATX network to produce suitable labels, then these labels are given to the AC network as a target along with the inputs to make a suitable trading decision. The AC network structure consists of four layers consisting of 1 LSTM layer with 128 neurons and 3 FC layers with 32, 64, and 64 neurons respectively. In this structure, the output of the fourth layer is provided to the Actor and Critic to make the appropriate trading decision and predict the input data label. In the proposed structure, the Actor has 2 heads, one head performs the trading decisions (buying, selling, nothing) and the other head is related to the auxiliary task that checks the labels of the input data and corrects the action. The Critic network is also used to predict the expected reward and evaluate the Actor's performance to help improve the Actor's performance and improve its predictions. The basic DRL algorithm used in this research is the PPO, whose performance has been improved by using auxiliary tasks. Therefore, the proposed structure is called PPO+AXT.

### D. Evaluation Metrics

The following three criteria have been used to evaluate the performance of models.

#### 1) The Amount of Cumulative Profit

As mentioned earlier, the agent is rewarded by the environment for acting as each time step. In this problem, this main reward is called PnL, which is calculated in time step t using:

$$Total\ Return = \sum_t r_t^{PnL} \ . \ r_t^{PnL} = \{-1.0. +1\} z_t \qquad (11)$$

Where, $\{-1, 0, 1\}$ are trading actions that indicate selling, doing nothing (stay out of market), and buying respectively, and $z_t$ is the return. The resulting cumulative profit is the sum of the profits of all time steps.

#### 2) Sharp Ratio

This criterion is used to evaluate the risk of trading and is calculated as follows:

$$Sharp\ Ratio = (returns\ mean) / (std\ of\ returns) \quad (12)$$

Here, *std* indicate standard deviations of returns

**3) The Percentage of Performance Improvement**

Since the conditions and training environment of our work are not the same as the basic article [3], the above parameters cannot be used to compare the two works. To compare the proposed model with the model of the basic article, the percentage of performance improvement (PPI) criterion is used, which is calculated as follows:

$$PPI = \frac{(value_{new} - value_{original})}{|value_{original}|} \times 100 \quad (13)$$

## IV. RESULTS

In this section, after introducing the used datasets, the mechanism of setting the parameters of the environment and tuning meta-parameters, the results of the proposed model will be presented and compared with the basic PPO model.

### A. Datasets

In this article, two separate datasets are used. The first dataset (DS1) is similar to [3] and the second dataset (DS2) is used from the data of recent years. The reason for using two separate datasets is that the DS1 was collected during the golden period of the European and American economy, and the DS2 was collected during the war in Ukraine, the coronavirus, and the economic crisis in Europe and the United States, so that we can compare our model performance in two completely different periods in the field of trading markets.

**DS1:** This dataset contains forex market data for the EUR/USD currency pair from the beginning of 2009 to 07/31/2017 for 1-hour candles, extracted using the MetaTrader 5 terminal. The data from 2009 to the end of 2016 were used for training and from the first 7 months of 2017 for backtesting.

**DS2:** This dataset contains forex market data for the EUR/USD currency pair from 03/03/2013 to 08/01/2023 for 1-hour candles, extracted using the MetaTrader 5 terminal. Data from 2013 to the end of 2022 were used for training and data from the first 7 months of 2023 for backtesting.

### B. Parameters and Hyper Parameters Tuning

In this research, the OPTUNA framework [24] was used to obtain the optimal value of meta-parameters of AE and K-Means algorithms. Using OPTUNA for AE, the best batch size, Learning Rate, and Encoding Dimension are 32, 0.0000879678, and 12, respectively, and 12 clusters are suggested for the K-Means algorithm.

In the environment of this problem, every 600-time steps are considered as an Episode, and 1000000 time steps are considered for DRL agent training. The value of epsilon (ε) in the proposed algorithm is set to 0.2.

| | PPO | PPO+AXT | PPI% |
|---|---|---|---|
| **Overall Return** | -25.2% | 14.86 % | 158.8% |
| **Sharp-Ratio** | -2.618 | 0.249 | 109.2% |

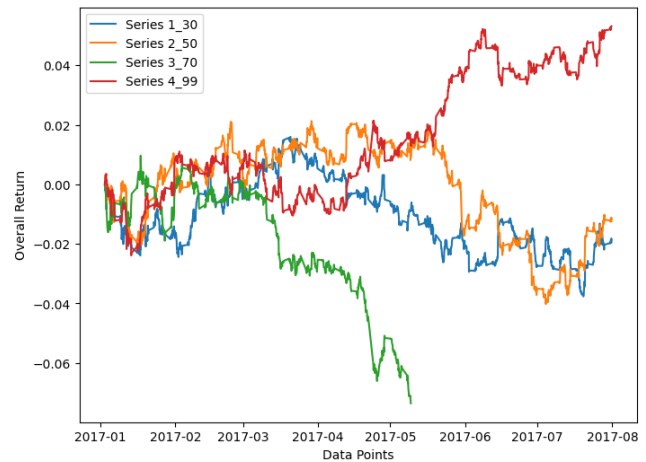

**Fig. 3**. PPO model Backtesting Performance on DS1

### C. Model Backtesting Performance

To evaluate the performance of the proposed model and to trust the obtained results, the model training process has been carried out based on different seeds, which are considered 30, 50, 70, and 99 respectively, and the average result of 4 different run was considered as models overall performance. The number provided after the name of the series in Fig. 3 to 6 indicated the seed number. Table I shows the average performance of the basic PPO model and the proposed model on the DS1 in the time step of 1000000. The results show that the PPO+AXT has been able to obtain promising results compared to the agent without an auxiliary task and has proven the assumption that the use of an auxiliary task improves learning in the reinforcement learning agent. The proposed model shows a better average performance than the base PPO model.

Figure 3 shows the return performance of the PPO agent in the 7-month backtesting period on DS1 for 4 different seeds. As it is clear, the basic PPO agent for the seed value of 70 did not have a good return and it was called, and by the seed value of 30 and 50 the agent did not earn a good return performance, but the execution by the seed value of 90 obtained a good return.

Figure 4 shows the performance of the PPO+AXT agent in the 7-month backtesting period on DS2 for 4 different seeds. As it is clear, the PPO+AXT agent for a seed value of 30 did not perform well and with a seed value of 50, it worked well at first, but then its efficiency decreased. The proposed model with seed values of 99 and 70 has achieved good results.

TABLE I
Backtesting Average Performance on DS1

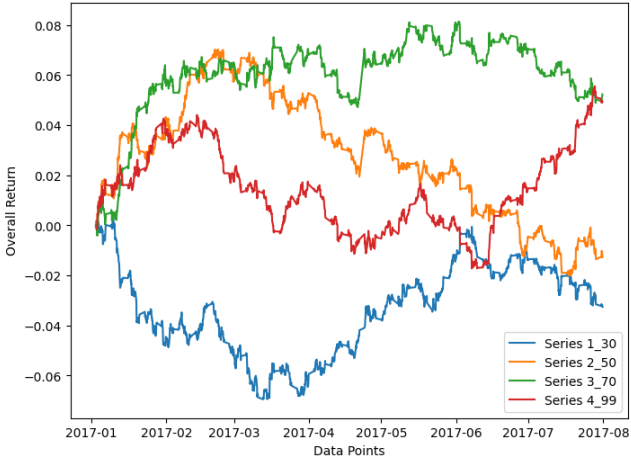

**Fig. 4.** PPO+AXT Backtesting Performance on DS1

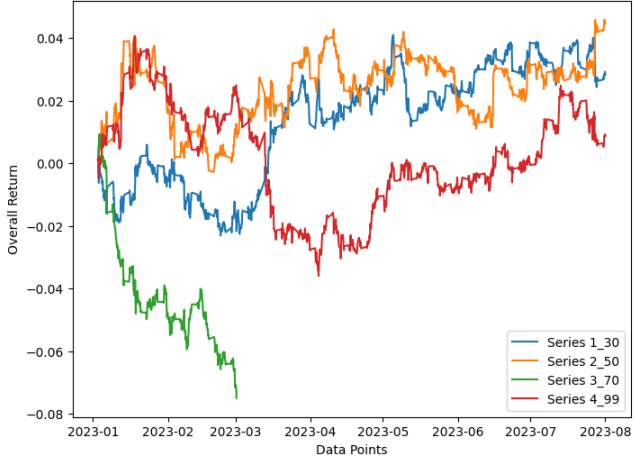

**Fig. 5**. PPO model Backtesting Performance on DS2

Table II shows the backtesting results of the basic model and the proposed model on the DS2 at a time step of 1,000,000. The obtained results show that the proposed PPO+AXT model has been able to perform much better than the basic PPO model and has a significant improvement.

Figure 5 shows the return performance of the PPO agent in the 7-month backtesting period on Ds2 for 4 different seeds. As it is clear, the PPO agent for the seed value of 70 did not have a good return and it was called. The seed value of 99 had a good return at first and then its return decreased. But, seeds with the value of 30 and 50 have achieved good results.

Figure 6 shows the performance of the PPO+AXT agent in the 7-month backtesting period on DS2 for 4 different seeds. As it is clear, the basic reinforcement learning agent for seeds with a value of 30 did not have a good return, but it achieved a good return for the other 3 seeds.

## V. DISCUSSION

We have compared and evaluated the results of this article, the use of auxiliary tasks, with the results of the basic article, that used additional rewards [3]. In the article [3], the model without additional reward is calculated through the DQN algorithm, and the model with additional reward is calculated

### TABLE II
### Backtesting Average Performance on DS2

|  | **PPO** | **PPO+AXT** | **PPI %** |
|---|---|---|---|
| **Overall Return** | 2.123% | 42.228% | 1891.51% |
| **Sharp-Ratio** | -2.933 | 0.473 | 116.4 % |

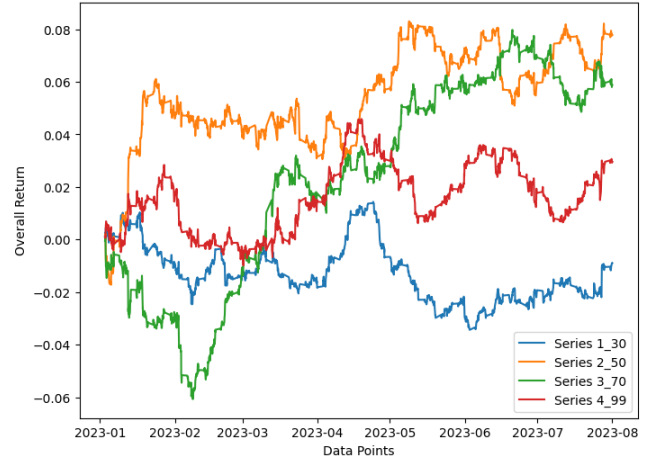

**Fig. 6.** PPO+AXT Backtesting Performance on DS2

### TABLE III
### Performance of PPO + Additional reward model [3]

|  | **DQN [3]** | **PPO+Trail[3]** | **PPI %** |
|---|---|---|---|
| **Return** | 3.3% | 6.2% | 87.88% |
| **Sharp Ratio** | 0.753 | 1.525 | 102.67% |

### TABLE IV
### PPI% of the Proposed Model Vs. Model [3]

|  | **PPO+AXT** | **PPO+Trail [3]** |
|---|---|---|
| **Return** | 158.85% | 87.88% |
| **Sharp Ratio** | 109.20% | 102.67% |

through the PPO algorithm. Table III shows that the PPO+Trail agent has shown better performance than the DQN agent and the additional reward problem has been able to improve the performance of the PPO agent [3].

Table IV shows the PPI% of the model of the article [3] with the PPI% of the proposed model on DS1. The results have shown that in all criteria, the proposed model has shown better performance. The results indicated that the combination of the PPO algorithm with auxiliary work is better than its combination with additional reward and has provided better performance.

In this research, to provide standard results, the model was trained and evaluated with different seeds, respectively, with values of 30, 50, 70, and 99. The reason for using this method is that random seeds are generated every time through the seed method, and these random seeds tell the agent from which state to start learning, if we start the training again, it will be from the same training state and makes the results to be reproducible.

By calculating the performance improvement percentage of the model [3] and the proposed model in this study, for both the DS1 and DS2, it can be concluded that the proposed model of this research has significantly increased in both return and sharp-ratio performance criteria. The percentage of performance improvement in the Overall Return criterion is 158.85%, 1891.51%, and 87.88% respectively for the PPO, the PPO+AXT, and [3], which shows a significant improvement in the performance of the proposed model in this research. In terms of Sharp-Ratio, the percentage of performance improvement has been obtained as 109.20%, 116.04%, and 102.67% respectively, which shows that the trading risk has been reduced in the proposed model and the trader can trust the model. The presented results showed the powerfulness of the proposed PPO+AXT model.

## VI. CONCLUSIONS

To improve the performance of the deep reinforcement learning agent in the forex market, increase the trading profit, and reduce the investment risk, a new model using the PPO algorithm is proposed. This algorithm uses the Actor-Critic structure to monitor the trading environment and make optimal trades. To better monitor the trading environment, correctly evaluate inputs, and find suitable trading positions in the proposed model, a new function has been used as an auxiliary task. The new auxiliary task uses AE and K-Means algorithm to cluster the input data and consider the cluster as the input data label. The Actor network has tried to adapt to the environmental conditions by predicting the input data label and comparing them with the output of auxiliary task, and in this way, by increasing its knowledge, it has improved its trading strategy to increase its accuracy. The performance of the proposed model using the auxiliary task compared to the basic model without the auxiliary task and comparing it with the paper [3] shows a significant improvement, which shows the efficiency of the proposed method. The results of applying the proposed model on two separate datasets have confirmed the main idea of this research about shaping rewards using auxiliary tasks. Therefore, labeling the input data using the unsupervised method, makes it possible for the trading agent to earn more knowledge of the environment and, with increased awareness, find a more optimal strategy for trading in the complex and dynamic forex market. As a result, the proposed agent model could be used for trading in this market.